\documentclass[prd,showpacs,preprintnumbers]{revtex4}
\usepackage[mathscr]{eucal}
\usepackage{color}
\usepackage{amsmath}
\usepackage[dvips]{graphicx}
\usepackage{epsf}
\usepackage{bm}
\newcommand{\be}{\begin{equation}}

\newcommand{\ee}{\end{equation}}
\newcommand{\bea}{\begin{eqnarray}}
\newcommand{\eea}{\end{eqnarray}}

\def\slr#1{\setbox0=\hbox{$#1$}           % set a box for #1
   \dimen0=\wd0                                 % and get its size
   \setbox1=\hbox{/} \dimen1=\wd1               % get size of /
   \ifdim\dimen0>\dimen1                        % #1 is bigger
      \rlap{\hbox to \dimen0{\hfil/\hfil}}      % so center / in box
      #1                                        % and print #1
   \else                                        % / is bigger
      \rlap{\hbox to \dimen1{\hfil$#1$\hfil}}   % so center #1
      /                                         % and print /
   \fi}

\def\be{\begin{eqnarray}}
\def\ee{\end{eqnarray}}

\begin{document}

\preprint{BARI-TH 515/05}

\title{Ginzburg-Landau approach to the three flavor LOFF phase of QCD}
\author{R. Casalbuoni}
\affiliation{Dipartimento di Fisica, Universit\`a di Firenze,
I-50019 Firenze, Italia} \affiliation{I.N.F.N., Sezione di Firenze,
I-50019 Firenze, Italia}
\author{R. Gatto}
\affiliation{D\'epart. de Physique Th\'eorique, Universit\'e de
Gen\`eve, CH-1211 Gen\`eve 4, Suisse}
\author{N. Ippolito}
\author{G. Nardulli}
\author{M. Ruggieri}
\affiliation{Universit\`a di Bari, I-70126 Bari, Italia}
\affiliation{I.N.F.N., Sezione di Bari, I-70126 Bari, Italia}
\date{\today}

\begin{abstract}
We explore, using a Ginzburg-Landau expansion of the free energy,
the Larkin-Ovchinnikov-Fulde-Ferrell (LOFF) phase of QCD with three
flavors, using the NJL four-fermion coupling to mimic gluon
interactions. We find that, below the point where the QCD
homogeneous superconductive phases  should give way to the normal
phase, Cooper condensation of the pairs $u-s$ and $d-u$ is possible,
but in the form of the inhomogeneous LOFF pairing.
\end{abstract}


\maketitle

\section{Introduction}

At high quark density and small temperatures Quantum-Chromo-Dynamics
(QCD) predicts Cooper pairing of quarks due to the existence of an
attractive quark interaction in the color antisymmetric channel,
see~\cite{barrois,alford,rapp} and for reviews
\cite{review,Nardulli:2002ma}. At extreme densities the
energetically favored phase is the Color-Flavor-Locking (CFL) phase,
characterized by a spin 0 diquark condensate antisymmetric in both
color and flavor \cite{alford2}; at intermediate densities the
situation is much more involved, because one cannot neglect the
strange quark mass and the differences $\delta\mu$ in the quark
chemical potentials induced by $\beta$ equilibrium. Several ground
states have been considered in the literature, from the 2SC
phase~\cite{alford}, to the gapless phases
g2SC~\cite{Shovkovy:2003uu} and gCFL
\cite{Alford:2003fq,Alford:2004hz}. The gapless phases  are instable
as shown by imaginary gluon Meissner masses (for g2SC
see~\cite{Huang:2004bg}, for gCFL see~\cite{Casalbuoni:2004tb}
and~\cite{Fukushima:2005cm}). This
 seems to be connected to the existence of
gapless modes in these phases \cite{Alford:2005qw}. An instability
is present also in the 2SC phase~\cite{Huang:2004bg}. Though this
phase has no gapless mode, imaginary gluon masses are present when
the gap $\Delta$ and $\delta\mu$ satisfy the condition
$\Delta/\sqrt{2}\le\delta\mu\le\Delta$.

Another superconductive state discussed in the literature is the
Larkin-Ovchinnikov-Fulde-Ferrell (LOFF) \cite{LOFF2} phase. The
relevance of this phase is based on the possibility that, for
appropriate values of $\delta\mu$, it can be advantageous for quarks
to form pairs with non-vanishing total momentum: ${\bf p_1}+{\bf
p_2}=2{\bf q}\neq 0$, see \cite{Alford:2000ze,Bowers:2002xr} and for
a review \cite{Casalbuoni:2003wh}. As far as instability is
concerned, the authors in ~\cite{Giannakis:2004pf} have shown that,
with  two flavors, the instability of 2SC implies that the LOFF
phase is energetically favored. Moreover, in the LOFF phase with two
flavors the gluon Meissner masses are real~\cite{Giannakis:2005vw}.

Thus far only the case of two species for the LOFF phase has been
studied.  This is not justified in QCD. At intermediate densities
all the three quarks: $u$, $d$ and $s$ should be considered. The
three flavor problem is however much more involved and difficult to
work out. We present here a first attempt to study the three flavor
LOFF phase of QCD. Our approach is  based on a Ginzburg-Landau (GL)
expansion of the free energy. Differently from the CFL phase, where
quark matter is in $\beta$ equilibrium while being also electrically
and color neutral, here we should impose these conditions. We shall
consider in the sequel only $\beta$-equilibrated and electrically
neutral quark matter, while assuming that the color-chemical
potentials vanish. This is an approximation we discuss below.

\section{Gap equation}

To get the gap equation in the Ginzburg Landau approximation, we
start with the Lagrangean density for three flavor ungapped quarks:
\begin{equation}
{\cal L}=\bar{\psi}_{i\alpha}\,\left(i\,D\!\!\!\!
/^{\,\,\alpha\beta}_{\,\,ij} -M_{ij}^{\alpha\beta}+
\mu^{\alpha\beta}_{ij} \,\gamma_0\right)\,\psi_{\beta j}
\label{lagr1}\ .
\end{equation}
 $M_{ij}^{\alpha\beta} =\delta^{\alpha\beta}\, {\rm diag}(0,0,M_s)
$ is the mass matrix and
$D^{\alpha\beta}_{ij}=\partial\delta^{\alpha\beta}\delta_{ij}+
igA_aT_a^{\alpha\beta}\delta_{ij}$; $\mu_{\alpha\beta}^{ij}$ is a
diagonal color-flavor matrix depending in general on $\mu$ (the
average quark chemical potential), $\mu_e$ (the electron chemical
potential), and $\mu_3,\,\mu_8$, related to color
\cite{Alford:2003fq}. We do not require color neutrality and we work
in the approximation $\mu_3=\mu_8=0$, which might be justified by
the results of Ref. \cite{Alford:2003fq} for the gCFL phase showing
that  $\mu_3$ and $\mu_8$ assume in general small values (at least
in the region of interest, see later). Therefore in this paper
\begin{equation}\mu^{\alpha\beta}_{ij}=(\mu\delta_{ij}-\mu_e
Q_{ij})\delta^{\alpha\beta}=\mu_{i}\,\delta_{ij}\delta^{\alpha\beta}\
,
\end{equation} where $Q$ is the quark electric-charge matrix.

We treat the strange quark mass at  the leading order in the $1/\mu$
expansion; this corresponds to a shift in the chemical potential of
the $s$ quark:
$\displaystyle\mu_s\rightarrow\mu_s-\frac{M_s^2}{2\mu}$. This is the
same approximation used in Refs.~\cite{Alford:2003fq},
\cite{Casalbuoni:2004tb} for the study of the gCFL phase. Therefore:
  \be \mu_u=\mu-\frac{2}{3}\mu_e~,\ \mu_d=\mu+\frac{1}{3}\mu_e~,\
\mu_s=\mu+\frac{1}{3}\mu_e-\frac{M_s^2}{2\mu}~.\label{eq:defChemPotQuarks}\ee

Another approximation we employ is the  High Density Effective
Theory (HDET), see \cite{Hong:1998tn,Beane:2000ms,Casalbuoni:2003cs}
and, for a review, \cite{Nardulli:2002ma}. Here one decomposes the
quark momentum  into a large component, proportional to $\mu$, and a
residual small component: $ {\bf p}=\mu{\bf n}+{\bf \ell}$; ${\bf n
}$ is a unit vector and $\ell$ is the small residual momentum.
Moreover one introduces $\bf n$-dependent fields  $\psi_{\bf n} $
and $\Psi_{\bf n}$ by the Fourier decomposition
\begin{equation}
\psi(x)=\int\frac{d{\bf n}}{4\pi}e^{i\,\mu{\bf n}\cdot{\bf
x}}\,\left(\psi_{\bf n}(x)+\Psi_{\bf n}(x)\right)\label{decomp} \,;
\end{equation}
$\psi_{\bf n}$ and $\Psi_{\bf n}$  correspond to positive and
negative energy solutions of the Dirac equation.

 Substituting the expression (\ref{decomp}) in the
Eq.~(\ref{lagr1}) one gets at the leading order in $1/\mu$
\begin{equation} {\cal L}= \int\frac{d{\bf n}}{4\pi}\,
\psi_{{\bf n},i\alpha}^\dagger\left(i V \cdot D^{\alpha\beta}_{ij} +
\bar\mu_i\delta^{\alpha\beta}\delta_{ij} \right) \psi_{{\bf n},\beta
j} \label{L11}\, ,
\end{equation}
where $V^\mu=(1,{\bf n}),~\tilde{V}^\mu=(1,-{\bf n})$ and $\bar\mu_i
= \mu_i-\mu$.

It is convenient to change the basis for the spinor fields by
defining $\psi_A = (\psi_{ur},
 \psi_{dg},\psi_{bs},\psi_{dr},\psi_{ug},
 \psi_{sr},\psi_{ub},\psi_{sg},\psi_{db})$.
This change of basis is performed by unitary matrices $F_A$, whose
explicit expression can be found in Ref. \cite{Casalbuoni:2004tb}.
To the Lagrangean in Eq. (\ref{L11}) we add a Nambu-Jona Lasinio
four fermion coupling treated in the mean field approximation.
This corresponds to the same coupling and the same approximation
used in Ref. \cite{Alford:2004hz}. The gap term in the resulting
Lagrangean is conveniently treated by introducing the Nambu-Gorkov
field
\begin{equation}
\chi_A=\frac{1}{\sqrt{2}}\left(\begin{array}{c}
  \psi_{\bf n} \\
  C\,\psi^*_{- \bf n}
\end{array}\right)_A \label{nambu-gorkov}
\end{equation}
so that the Lagrangean reads
\begin{equation}
L=\frac{1}{2}\sum_{A,B}\int\frac{d{\bf
n}}{4\pi}\,\int\frac{dE\,d\xi}{(2\pi)^2}\,\chi^\dagger_A\,\left(
\begin{array}{cc}
  \left(E-\xi + \bar\mu_{A}\right)\delta_{AB} & -\Delta_{AB}({\bf r}) \\
  -\Delta_{AB}^*({\bf r}) &  \left(E+\xi - \bar\mu_{A} \right)\delta_{AB}
\end{array}
\right)\,\chi_B \label{kinetic}
\end{equation}
where $E$ is the energy, $\xi\equiv {\bf \ell}\cdot{\bf n}$ is the
component of the residual momentum along $\bf n$ and satisfies:
$|\xi|<\delta$, with $\delta$ an ultraviolet cutoff. Moreover $
(\bar\mu)_{A} \ = \ \left(\bar\mu_{u}, \bar\mu_{d}, \bar\mu_{s},
\bar\mu_{d}, \bar\mu_{u}, \bar\mu_{s}, \bar\mu_{u}, \bar\mu_{s},
\bar\mu_{d}\right)$.

We assume the pairing ansatz
\begin{equation}
<\psi_{i\alpha}\,C\,\gamma_5\,\psi_{\beta j}> =
\sum_{I=1}^{3}\,\Delta_I({\bf r})\,\epsilon^{\alpha\beta
I}\,\epsilon_{ijI}~\label{cond}
\end{equation}with \be \Delta_I ({\bf r}) = \Delta_I
\exp\left(2i{\bf q_I}\cdot{\bf r}\right)~. \label{eq:1Ws}\ee In
other words, for each inhomogeneous pairing we assume a
Fulde-Ferrell ansatz; $2{\bf q_I}$ represents the  momentum of the
Cooper pair. The gap matrix $\Delta_{AB}$ in (\ref{kinetic}) can be
expressed in terms of the three independent functions
$\Delta_{1}({\bf r})$, $\Delta_{2}({\bf r})$, $\Delta_{3}({\bf r})$
describing respectively $d-s$, $u-s$ and $u-d$ pairing. The explicit
expression of $\Delta_{AB}$ can be found in \cite{Alford:2003fq},
\cite{Casalbuoni:2004tb}.

To write down the gap equation it is useful to introduce the
following components of the free quark propagator \be
\left[S_0^{11}\right]_{AB}\equiv\frac{\delta^{AB}}{E-\xi +
\bar\mu_{A}}~,~~~~~
 \left[S_0^{22}\right]_{AB}\equiv\frac{\delta^{AB}}{E+\xi - \bar\mu_{A}}~.
\label{freeQuarksProp}\ee The quark propagator is the matrix \be
S_{AB}=\left(
\begin{array}{cc}
  S_{11} & S_{12} \\
  S_{21} & S_{22}
\end{array}  \right)_{AB}  \label{quarkPropmatr} \ee
whose components satisfy the Gorkov equations \be S_{11}=S_0^{11} +
S_0^{11}\Delta({\bf r})S_{21}~,~~~~~S_{21}=S_0^{22}\Delta^*({\bf
r})S_{11}~. \label{eq:gorkov} \ee $S_{21}$ is the anomalous
propagator involved in the gap equation.

The wave vectors ${\bf q_I}$ should be derived by minimizing the
free energy. We will fix the norms $|{\bf q_I}|$ by a minimization
procedure. As to their directions, we will limit the analysis to
 four structures, choosing among them the one with the smallest value of the
energy.
 The first structure has  all ${\bf q_I}$ along
the positive $z-$axis. The structures 2, 3, 4 have, respectively,
${\bf q_1}$, ${\bf q_2}$, ${\bf q_3}$ along the positive $z-$axis
(the remaining two momenta along the negative $z-$axis). This is
obviously a limitation. It is justified by our final results that
show the existence of a range of values of the strange quark mass
where the LOFF phase, even with these limitations, is favored in
comparison with other QCD phases.

The gap equation in the HDET formalism can be written as follows
\cite{Nardulli:2002ma}
\begin{equation}
\Delta^*_{AB}({\bf r}) =\,i\,3G\,V^\mu\tilde{V}^\nu
\sum_{C,D=1}^9h^*_{AaC}h_{DbB}\int\frac{d\,{\bf n}}{4\pi}
\int\frac{d^3\,\ell}{(2\pi)^3}
\int\frac{dE}{2\pi}\,S_{21}(E,\ell)_{CD}\,g_{\mu\nu}\,\delta_{ab}~,
\label{eq:GapEq122}
\end{equation}
where $S_{21}$ is given in Eq. \eqref{eq:gorkov}; in the above
equation $h_{DbB}$ is a Clebsch-Gordan coefficient. It is expressed
by the formula  $h_{DbB}={\text{Tr}}[F_D^\dagger T_b F_B]$  in terms
of the unitary matrices $F_A$ used to write the quark fields as in
(\ref{nambu-gorkov}), i.e. in the basis $A=1,\cdots, 9$. $G$ is the
Nambu-Jona Lasinio coupling constant, of dimension mass$^{-2}$. In
what follows, we shall get rid of $G$ introducing the value of the
CFL gap parameter $\Delta_0$ as a measure of the strength of the
interaction (see below, Eq.~(\ref{def:newUltra})).

\section{Ginzburg-Landau expansion}

Performing the Ginzburg-Landau expansion of the propagator \be
S_{21}=S_0^{22}\Delta^*S_0^{11} +S_0^{22}\Delta^*S_0^{11}\Delta
S_0^{22}\Delta^*S_0^{11} +O(\Delta^5)\, \label{eq:GLexpansion} \ee
we get
\begin{equation}
\Delta_I=\Pi_I\,\Delta_I ~+~\sum_{J} J_{IJ}\,\Delta_I\,\Delta_J^2
~+~ O(\Delta^5)~,~~~~~I=1,2,3~.\label{eq:GapEquation11}
\end{equation}
Let us comment on the functions $\Pi_I$ and $J_{IJ}$ appearing in
this expansion. $\Pi_I$ are defined as follows:
$\Pi_1=\Pi(q_1,\delta\mu_{ds})~,$~
$\Pi_2=\Pi(q_2,\delta\mu_{us})~,$~$\Pi_3=\Pi(q_3,\delta\mu_{ud})~,$
with \be
\delta\mu_{ud}\equiv\frac{\bar\mu_d-\bar\mu_u}{2}=\frac{\mu_e}{2}~,~~~~~
\delta\mu_{us}\equiv\frac{\bar\mu_s-\bar\mu_u}{2}=\frac{\mu_e}{2}-\frac{M_s^2}{4\mu}~,~~~~~
\delta\mu_{ds}\equiv\frac{\bar\mu_s-\bar\mu_d}{2}=-\frac{M_s^2}{4\mu}~.
\label{deltamudefini}\ee and \be \Pi(q,\delta\mu)=
1~+~\frac{2G\mu^2}{\pi^2} \left(1 -
\frac{\delta\mu}{2q}\log\left|\frac{q+\delta\mu}{q-\delta\mu}\right|
- \frac{1}{2}\log\left|\frac{4(q^2
-\delta\mu^2)}{\Delta_0^2}\right|\right)\ .  \ee We note that $\Pi$
is analogous to the function determining the behavior of the free
energy in the GL approximation of the LOFF phase with two flavors.
We have introduced the parameter $\Delta_0$ to get rid of the
ultraviolet cutoff $\delta$. It is defined by
\begin{equation}
\Delta_0\equiv2\delta\,\exp\left\{-\frac{\pi^2}{2\,G\mu^2}\right\}~.
\label{def:newUltra}
\end{equation}
 $\Delta_0$ is equal to the CFL gap for $M_s=0$ and
$\mu_e=0$ in the weak coupling limit, with no sextet condensation.
As for $J_{IJ}$, we have, for the diagonal components: $J_{11}\equiv
J_1\equiv J(q_1,\delta\mu_{ds})$, $J_{22}\equiv J_2\equiv
J(q_2,\delta\mu_{us})$, $J_{33}\equiv J_3\equiv
J(q_3,\delta\mu_{ud})$, with \be
J(q,\delta\mu)=-\frac{G\mu^2}{2\pi^2}\frac{1}{q^2-\delta\mu^2} ~ \
.\ee The off-diagonal term $J_{12}$ is
\begin{equation}
J_{12}=\frac{G\mu^2}{\pi^2} \int\frac{d{\bf
n}}{4\pi}\,\frac{1}{(2{\bf q_1}\cdot{\bf
n}+\mu_s-\mu_d-i\epsilon)\,(2{\bf q_2}\cdot{\bf
n}+\mu_s-\mu_u-i\epsilon)}~;\label{j12}\end{equation} $J_{13}$ is
obtained from $J_{12}$ in (\ref{j12}) by the exchange ${\bf q_2\to
q_3}$ and $\mu_s\leftrightarrow\mu_d$; $J_{23}$ from $J_{12}$ by
${\bf q_1\to q_3}$ and $\mu_s\leftrightarrow\mu_u$.

\section{Free energy}

Let us now consider the free energy $\Omega$. It is obtained by
integrating the gap equation. The result is
\begin{equation}
\Omega =\Omega_n+ \sum_{I=1}^3\left(\frac{\alpha_I}{2}\,\Delta_I^2
~+~ \frac{\beta_I}{4}\,\Delta_I^4 ~+~ \sum_{J\neq
I}\frac{\beta_{IJ}}{4}\,\Delta_I^2\Delta_J^2 \right) ~+~ O(\Delta^6)
\label{eq:OmegaDelta11}
\end{equation}
with \begin{equation}
 \Omega_n = -\frac{3}{12\pi^2}\left(\mu_u^4+\mu_d^4+\mu_s^4\right) -
\frac{\mu_e^4}{12\pi^2}    \label{eq:OmegaNorm11222}
\end{equation}
where the chemical potentials for quarks are defined in
Eq.~(\ref{eq:defChemPotQuarks}) and the coefficients are given by
\begin{equation}
  \alpha_I=\frac{2\,(1-\Pi_I)}{G}~,~~~~~~ \beta_I=-\frac{2\,J_I}{G}~,~~~~~~
\beta_{IJ}=-\frac{2\,J_{IJ}}{G}~.   \label{eq:defCoeff11}
\end{equation}
Electric neutrality is obtained by imposing the condition
\begin{equation}
-\frac{\partial\Omega}{\partial\mu_e}=0~, \label{eq:electrNeutra111}
\end{equation}which, together with the gap equations, gives,
for each value of the strange quark mass, the electron chemical
potential $\mu_e$ and the gap parameters $\Delta_I$. Moreover one
should determine ${\bf q_I}$ by searching for the energetically
favored solution. This is a complex task as it would require the
simultaneous solution of the previous
equations~\eqref{eq:electrNeutra111} and~\eqref{eq:GapEquation11}
together with:
\begin{equation}
0~=\ \frac{\partial\Omega}{\partial q_I}=
\Delta_I\frac{\partial\alpha_I}{\partial q_I} +
\Delta_I\sum_{J=1}^{3}\Delta_J^2\frac{\partial\beta_{IJ}}{\partial
q_I}\ ,~~~~~I=1,2,3~. \label{eq:q}
\end{equation}
Moreover one should look for the most energetically favored
orientations of the three vectors ${\bf q_I}$ in space. A complete
analysis is postponed to a future paper; as discussed above we have
limited the analysis to the four structures characterized by all
vectors $\bf q_I$ parallel or antiparallel to the same axis. Even
with this limitation we are able to prove that there exists a window
of values of $M_s$ where the LOFF phase is favored in comparison
with other phases of QCD, as it will be seen below. As to the norms
$|{\bf q_I}|$, since we work in the GL approximation, we can neglect
the ${\cal O}(\Delta^2)$ terms in (\ref{eq:q}). As a consequence we
simply get $\displaystyle\frac{\partial\alpha_I}{\partial q_I}=0 $,
which, being identical to the condition for two flavors, gives the
result $q_I=1.1997|\delta\mu_I|$ \cite{LOFF2,Alford:2000ze}.

\section{Results and discussion}

\begin{figure}[ht!]
\begin{center}
{\includegraphics[width=9.5cm]{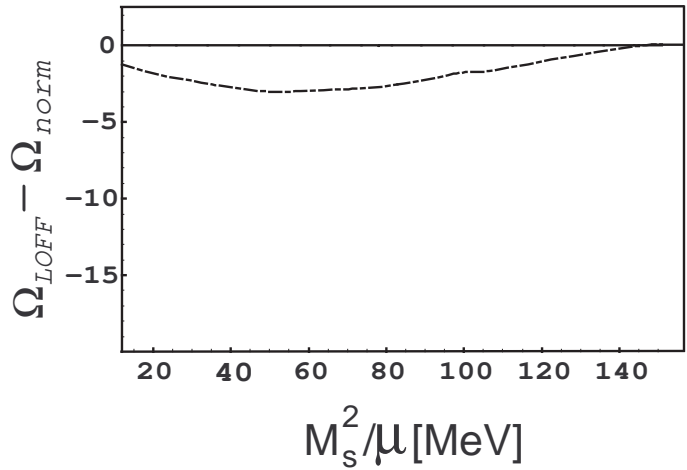}}
\end{center}
\caption{\label{FIG:omega500} Free energy difference
$\Omega_{LOFF}-\Omega_{norm}$ in units of $10^{6}$ MeV$^4$ plotted
versus  $M_s^2/\mu$ (in MeV). The result is obtained for $\mu=500$
MeV and $\Delta_0=25$ MeV. The line corresponds to the structures 1
and 2 with $\Delta_1=0,\Delta_2=\Delta_3$.}
\end{figure}

\begin{figure}[ht!]
\begin{center}
{\includegraphics[width=8.5cm]{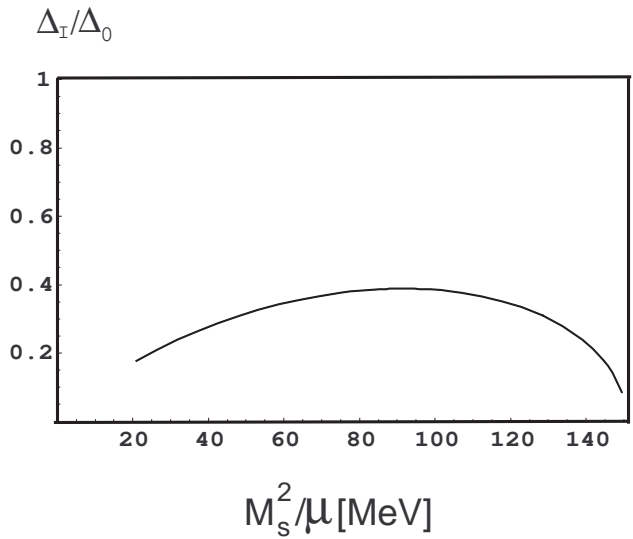}}
\end{center}
\caption{\label{FIG:gaps500} Gaps $\Delta_I/\Delta_0$ as functions
of $M_s^2/\mu$ (in MeV) for the structures 1 and 2; the curve
represents $\Delta_I=\Delta_2=\Delta_3$ whereas $\Delta_1=0$.}
\end{figure}

\begin{figure}[ht!]
\begin{center}
{\includegraphics[width=7cm]{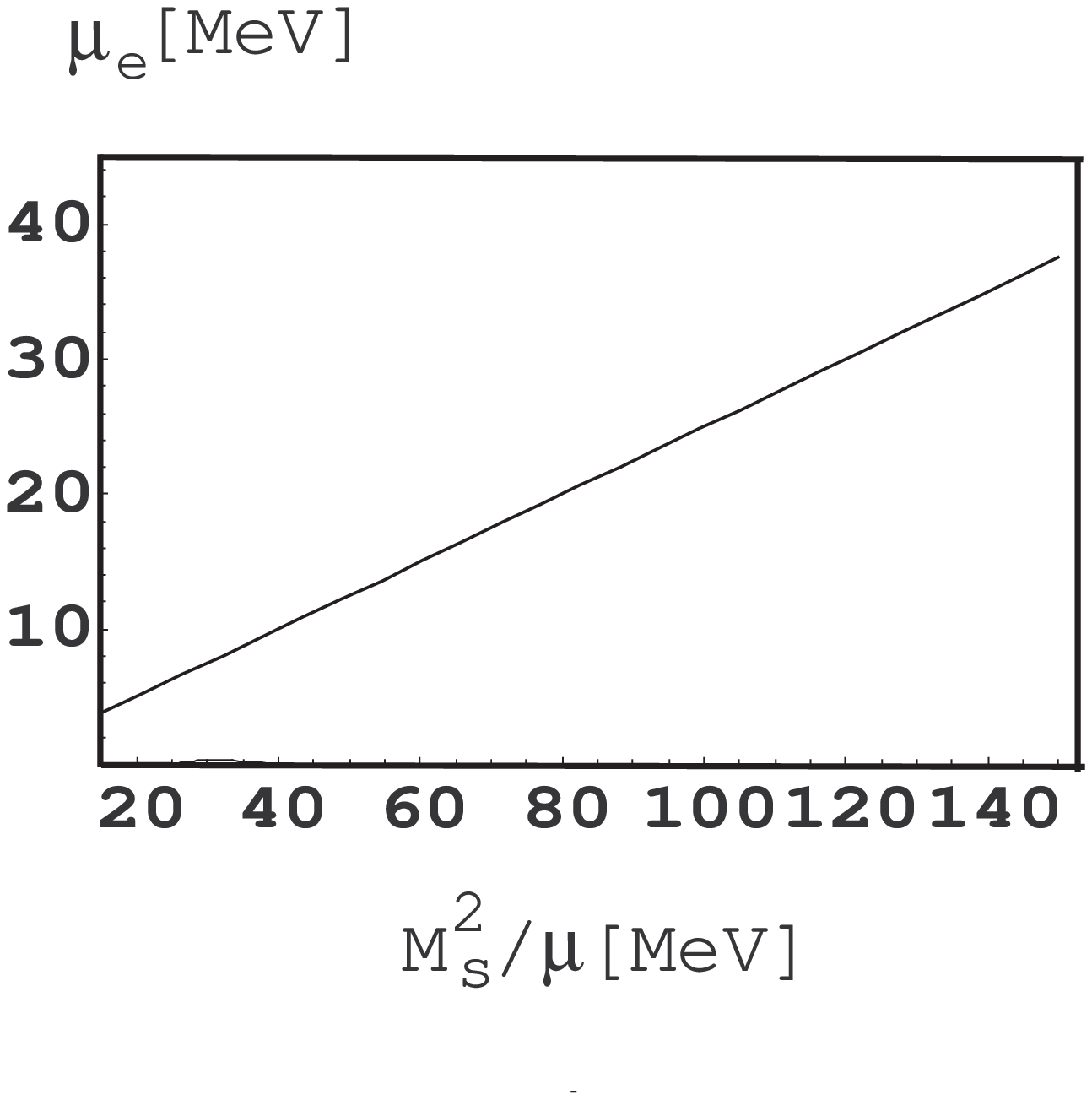}}
\end{center}
\caption{\label{FIG:mue500} The electron chemical potential $\mu_e$
as a function of  $M_s^2/\mu$. Units are MeV. The line corresponds
to $\Delta_2=\Delta_3$, $\Delta_1=0$. }
\end{figure}

\begin{figure}[ht!]
\begin{center}
{\includegraphics[width=14cm]{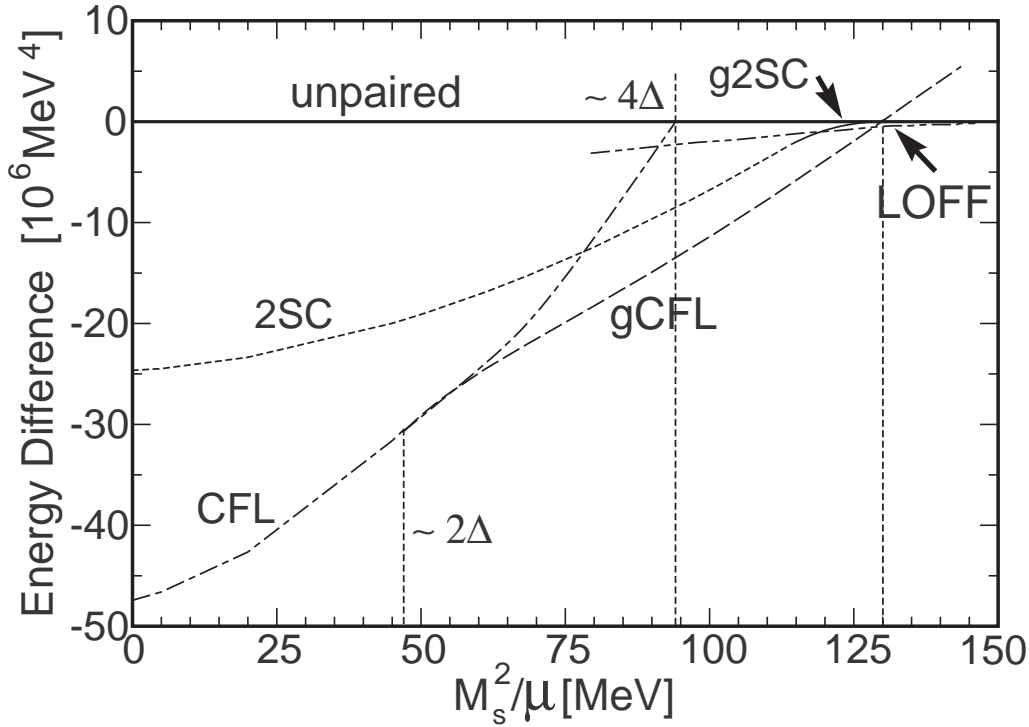}}
\end{center}
\caption{\label{FIG:comparison500} Free energy differences
$\Omega_{LOFF}-\Omega_{norm}$ in units of $10^{6}$ MeV$^4$ plotted
versus  $M_s^2/\mu$ (in MeV)  for various QCD phases.}
\end{figure}

Our results are summarized in Figs.
\ref{FIG:omega500}-\ref{FIG:comparison500}. In Fig.
\ref{FIG:omega500} we give $\Omega_{LOFF}-\Omega_{norm}$ (in units
$10^{6}$ MeV$^4$) as a function of $M_s^2/\mu$ (in MeV).

 We report only the energetically favored solution. It corresponds
 to $\Delta_1=0,\Delta_2=\Delta_3$ and
  $\bf q_2$, $\bf q_3$
   parallel (structure 1, which is identical in this case to the
   structure 2). We do not report the other solutions corresponding
    to local minima of $\Omega$. We also found a solution with $\Delta_1\neq 0$ and $\Delta_2
   \neq \Delta_3$ but it can only exist in a kinematical range where the LOFF
   phase is energetically disfavored in comparison with the CFL
   or the gCFL phases.
The structures 3 and 4 ($\bf q_2$, $\bf q_3$ antiparallel) have
almost vanishing gaps and correspond only to local minima of the
free energy. The results in this figure and in the subsequent ones
are obtained for $\mu=500$ MeV (for $\mu=400$ MeV the results are
qualitatively similar). The value of the CFL gap for zero strange
quark mass is fixed at $\Delta_0=25$ MeV. This is the same value
used in Ref. \cite{Alford:2004hz}. This choice, as well as the same
form of the NJL coupling, with the same approximation, allows a
comparison between our results and those of Ref.
\cite{Alford:2004hz}, see the discussion below.

In Fig. \ref{FIG:gaps500} we give  the gaps $\Delta_I/\Delta_0$ as
functions of  $M_s^2/\mu$ (in MeV). The line represents the solution
$\Delta_2=\Delta_3$ for the structures 1 and 2 (in this case
$\Delta_1=0$).

In Fig. \ref{FIG:mue500} we present results for the electron
chemical potential $\mu_e$. The line correspond to the energetically
favored solution (structure 1$\equiv $2; we have neglected terms
suppressed in the $1/\mu$ expansion, consistently with the HDET
scheme). A glance at eq. (\ref{deltamudefini}) shows that $\mu_e$ is
given by $\mu_e\approx M_s^2/(4\mu)$, which  corresponds to a
symmetric splitting of the $s$ and $d$ Fermi surfaces around the $u$
Fermi surface. Therefore in this kinematical region we have $us$ and
$du$ pairings, with
$${\bf p_u}+{\bf p_s}\,=\, 2\,{\bf q_2}\,,\hskip2cm {\bf p_u}+{\bf p_d} =
\,2\,{\bf q_3}\,=\,2\, {\bf q_2}\,.$$ The
 gaps $\Delta_2$ ($us$ pairing) and $\Delta_3$ ($ud$ pairing) have to be
almost equal since they depend
 only on the absolute values of the splittings, see Fig. \ref{FIG:gaps500}. Since the
 separation between the $d$ and $s$ Fermi surfaces is higher,
 one does not expect $ds$ pairing, which is confirmed by
the result $\Delta_1=0$ in the region where LOFF prevails.
\\
\\
In Fig. \ref{FIG:comparison500} we present comparison of different
phases of QCD. In order to comment this figure, let us start
assuming that all the other phases are stable, meaning that in some
way it is possible to cure the instability due to the imaginary
gluon masses. In this case, following the graph for decreasing
values of $M_s^2/\mu$, we see that at about $M_s^2/\mu=150$ MeV the
LOFF phase has a free energy lower than the normal one. This is a
second order transition as it can be seen from Fig.
\ref{FIG:gaps500}. Then the LOFF state is energetically favored till
the point where it meets the gCFL line at about $M_s^2/\mu= 128$
MeV. This is a first order transition since all the gaps are
different in the two phases (for the gCFL case, see
\cite{Alford:2003fq}). Then the system stays in the gCFL phase up to
$M_s^2/\mu\approx 48$ MeV where it turns into the CFL phase via a
second order transition (see~\cite{Alford:2003fq}).

However, if the gapless phases are unstable, then they should not be
considered, and the LOFF phase is the stable phase from
$M_s^2/\mu=150$ MeV down to about $M_s^2/\mu=90$~MeV where the LOFF
line meets the CFL line, with a first order transition (this can be
seen by comparing our gaps with the $\Delta_{CFL}\approx 23$~MeV at
this value of $M_s^2/\mu$).

We should also add that  at the moment it is still unknown if  the
LOFF phase with three flavors suffers of  chromo-magnetic
instabilities. This  problem is left to future investigations.

\section{Conclusion and outlook}

We have explored in the framework of the Ginzburg-Landau expansion
the LOFF phase of QCD with three flavors, using the NJL
four-fermion coupling to mimic the gluon interactions. We have
worked on the ansatz of a single plane wave behaviour for each
quark pairing, which is the simplest generalization of the gCFL
phase that takes into account the possibility of anisotropic
condensation.  We found that near the point where the CFL phase
should give way to the normal phase, Cooper condensation takes
place in the form of the LOFF pairing. Our analysis has some
limitations. First, we have assumed vanishing color chemical
potentials $\mu_3,\,\mu_8$. The results of Ref.
\cite{Alford:2004hz} show that in the region where the LOFF
 state dominates the color chemical potential have
 rather small values, in particular smaller than $\mu_e$. However
non vanishing values of $\mu_3$ and  $\mu_8$ are expected to
increase the LOFF free energy and therefore a more complete
calculation is necessary. Second, we have considered the three
possible momenta $\bf q_I$ along the same direction. Third, more
than one plane wave might be present in each condensate. Finally we
have treated the strange quark mass at its leading effect, i.e. by a
shift in its chemical potential, which is also an approximation
\cite{Casalbuoni:2003cs}. We plan to address all these issues by a
more refined study in the future. \vskip.2cm

{\em Acknowledgments}.

We wish to thank M.~Mannarelli for useful discussions and comments
and K. Rajagopal and R. Sharma for helpful correspondence.

\end{document}